\documentclass[showpacs,aps,epsfig,nofootinbib]{revtex4}
\usepackage{mathrsfs}

\usepackage{graphicx}
\newcommand{\ba}{\begin{eqnarray}} \newcommand{\ea}{\end{eqnarray}}
\usepackage{amsfonts}
\usepackage{epstopdf}
\usepackage{latexsym}
\usepackage{amssymb}
\usepackage{amssymb}

\usepackage[center]{subfigure}

\begin{document}

 \newcommand{\bq}{\begin{equation}}
 \newcommand{\eq}{\end{equation}}
 \newcommand{\bqn}{\begin{eqnarray}}
 \newcommand{\eqn}{\end{eqnarray}}
 \newcommand{\nb}{\nonumber}
 \newcommand{\lb}{\label}

\newcommand{\PRL}{Phys. Rev. Lett.}
\newcommand{\PL}{Phys. Lett.}
\newcommand{\PR}{Phys. Rev.}
\newcommand{\CQG}{Class. Quantum Grav.}

\title{Resonance of Gaussian electromagnetic field to the high frequency gravitational waves}
\author{Jin Li$^{*}$} ,\author{Lu Zhang$^{*}$}\author{Kai Lin$^{\dagger}$}\author{Hao Wen}
\email{cqstarv@hotmail.com}
\affiliation {$^{*}$ College of Physics, Chongqing University,Chongqing 401331, China, and}
\affiliation {State Key Laboratory of Theoretical Physics, Institute of Theoretical Physics, Chinese Academy of Sciences,
Beijing 100190, China.}
\affiliation {$^{\dagger}$Instituto de F\'isica e Qu\'imica,
Universidade Federal de Itajub\'a, MG, Brasil} \affiliation
{Instituto de F\'isica, Universidade de S\~ao Paulo, CP 66318,
05315-970, S\~ao Paulo, Brazil} \

\date{\today}

\begin{abstract}
We consider a Gaussian Beam (GB) resonant system for high frequency gravitational waves (HFGWs) detection. At present, we find the optimal signal strength in theory through setting the magnetic component of GB in a standard gaussian form. Under the synchro-resonance condition, we study the signal strength (i.e., transverse perturbative photon fluxes) from the relic HFGWs (predicted by ordinary inflationary model) and the braneworld HFGWs (from braneworld scenarios). Both of them would generate potentially detectable transverse perturbative photon fluxes (PPFs). Furthermore we find optimal system parameters and the relationship between frequency and effective width of energy fluxes accumulation.

\end{abstract}


\maketitle
With the development of gravitational wave detectors in LF and VLF, people also hope to make a breakthrough in high-frequency gravitational wave detection. HFGWs in microwave band carry much different information of the universe evolution, it is necessary to build high-frequency gravitational wave detectors. The detectors in such frequency band mostly base on the interaction between GWs and EM fields\cite{1971,1978,Grishchuk2003}. For the GWs with frequency $\sim 10^{9}\rm{Hz}$, a perturbation EM field will be generated from the interaction between HFGWs and a static magnetic field. Then the perturbation EM field with resonant GB background produces transverse perturbative photon flux $n^{(1)}_{x}$\cite{scheme1,shotnoise1,PRD89}, which will be explained in the following context. In ordinary inflationary model with specific inflation parameters ($r=0.22$,$\beta=-2.015$,$\alpha=0.005$),  the dimensionless amplitude of the relic GWs in the microwave band (relic HFGWs) ($\sim 10^{8}-10^{10}\rm{Hz}$) can reach up to $h\sim10^{-30}-10^{-32}$ \cite{zhang1,zhang2,zhang3}. Furthermore the latest work of Giovannini\cite{arxivref1} shows that if the standard $\Lambda$CDM scenario\cite{arxivref1ref2} with a tensor-to-scalar-ratio $r_{T}\sim0.2$ is complemented by a high-frequency component, the maximal signal will appear around GHz frequency band. Our previous results indicate that the strength of the signal is around $10^{2}\rm{s}^{-1}$\cite{EPJC}-\cite{grg}. In this paper, we find that the signal strength can be further enhanced through setting the standard Gaussian form on longitudinal magnetic component of GB, furthermore the signal is independent from frequency bandwidth. In extra dimensions model, near the Earth the HFGWs emitted from braneworld black hole with wider frequency band ($10^{8}-10^{14}\rm{Hz}$) \cite{extrabrane2},\cite{extrabrane3} would be quite powerful with an upper limit of the dimensionless amplitudes around $h\sim 10^{-21}-10^{-22}$, so the HFGWs from the braneworld might likely arouse much stronger photon fluxes.

As the GWs from celestial bodies, the metric of HFGWs can be written as a small perturbation $h_{\mu\nu}$ to the flat spacetime $\eta_{\mu\nu}$.
\begin{equation}
g_{\mu\nu}=\eta_{\mu\nu}+h_{\mu\nu},
\end{equation}
where $h_{\mu\nu}$ represents a small correction to the background metric tensor. Supposing the HFGWs come along the z axis in our coordinate system and considering the metric can be expressed as the following form in Cartesian coordinates:
\begin{equation}
h_{\mu\nu}=\left(
              \begin{array}{cccc}
                0 & 0 & 0 & 0 \\
                0 & h_{11} & h_{12} & 0 \\
                0 & h_{21} & h_{22} & 0 \\
                0 & 0 & 0 & 0 \\
              \end{array}
            \right),
\end{equation}
where $h_{11}=-h_{22}=A_{\oplus}\rm{exp}[i(k_{g}z-\omega_{g}t)]$, $h_{12}=h_{21}=\rm{i}A_{\otimes}\rm{exp}[i(k_{g}z-\omega_{g}t)]$. Here $A_{\oplus}$ and $A_{\otimes}$ denote the amplitude of the $\oplus$ polarization and $\otimes$ polarization in the laboratory frame respectively. The relic HFGWs and HFGWs predicted by braneworld scenarios have different amplitude and frequency band. The former is a stochastic signal but the latter is a coherent signal with discrete spectrum~\cite{extrabrane2} from a fixed source. Therefore, it is meaningful to make a comparison of signal from relic HFGWs and braneworld HFGWs.
\begin{figure}[htbp]
\label{scheme}
\centerline{\includegraphics[height=5cm]{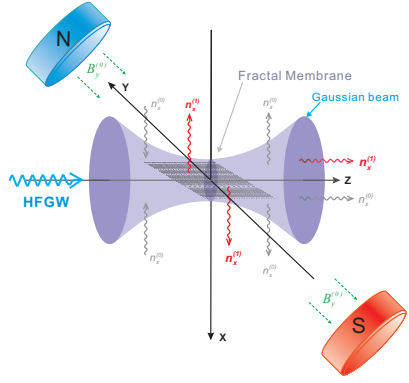}}
\caption{The configuration of the Gaussian Beam (GB) resonant system}
\end{figure}
In Figure 1, on y-axis the static magnetic field $\hat{B}^{(0)}_{y}$ would interact with the incoming HFGWs and generate the perturbative EM signals\cite{EPJC},\cite{PRD89}, which can be explained by the electro-dynamical equations in curved spacetime\cite{EPJC},\cite{PRD67},
\begin{equation}
\frac{1}{\sqrt{-g}}\frac{\partial}{\partial x^{\nu}}\left(\sqrt{-g}g^{\mu\alpha}g^{\nu\beta}F_{\alpha\beta}\right)=\mu_{0}J^{\mu},
\end{equation}
\begin{equation}
\nabla_{\alpha}F_{\mu\nu}+\nabla_{\nu}F_{\alpha\mu}+\nabla_{\mu}F_{\nu\alpha}=0,
\end{equation}
where $J^{\mu}$ indicates the four-dimensional electric current density. For the detector is immersed in vacuum, $J^{\mu}=0$. Using the perturbation methods ($F_{\mu\nu}=F^{(0)}_{\mu\nu}+\tilde{F}^{(1)}_{\mu\nu}+O(h^{2})$; $F^{(0)}_{\mu\nu}$ and $\tilde{F}^{(1)}_{\mu\nu}$ indicate the background and first-order perturbative EM tensor respectively) and neglecting the second and higher-order perturbation, the first-order perturbative EM fields along positive z-axis can be derived~\cite{EPJC}\cite{PRD67}\cite{PRD89}\cite{epjc28}. Here we only concern the first-order perturbative electric field $\tilde{E}_{y}^{(1)}$,
\begin{equation}
\label{perRelic}
\tilde{E}_{y}^{(1)}=-\frac{1}{2}A_{\otimes}\hat{B}^{(0)}_{y}k_{g}c(z+l_{1})\exp{[i(k_{g}z-\omega_{g}t)]}
\end{equation}
where the superscript $^{(0)},^{(1)}$ denote the quantities of background and perturbative field, respectively. And the notation $^{\sim}$ and $^{\wedge}$ represent the time-dependent and static value. The perturbative electric field $\tilde{E}_{y}^{(1)}$ is similar to a plane EM wave, and has accumulative effect in propagation direction. However in the laboratory scale, the accumulative effect in $z$ direction performs not so obvious on the transverse perturbative photon fluxes (PPFs).Certainly, if the background static magnetic field $\hat{B}^{(0)}_{y}$ is replaced by the background galactic-extragalactic magnetic fields, a significant accumulation effect would emerge \cite{PRD89}.

As our major part, a Gaussian Beam (GB) (see Fig.1) with a fundamental frequency mode\cite{epjc56} propagating along z-axis, which can provide a non-vanishing longitudinal magnetic field component $\tilde{B}^{(0)}_{z}$. Furthermore due to the equivalent status of electric and magnetic components in EM field, we assume the longitudinal magnetic field component in standard Gaussian form as
\begin{equation}
\tilde{B}^{(0)}_{z}=\frac{\psi_{0}}{\sqrt{1+(z/f)^{2}}}\exp\left(-\frac{r^{2}}{W^{2}}\right)\exp\left\{i\left(k_{e}z-\omega_{e}t-\arctan\frac{z}{f}+\frac{k_{e}r^{2}}{2R}+\delta\right)\right\}.
\end{equation}
and $\tilde{E}^{(0)}_{z}=0$, where $r^{2}=x^{2}+y^{2}$, $k_{e}=2\pi/\lambda$ (the wave
number), $f=\pi W^{2}_{0}/\lambda$,$W=W_{0}[1+(z/f)^{2}]^{1/2}$
($W_{0}$ is the minimum spot radius at $z=0$),$R=z+f^{2}/z$ (the
curvature radius of the wave fronts of the GB at $z$)
and $\delta$ is the phase difference between the GB and the resonant HFGWs. $\psi_{0}$ is the amplitude of the magnetic element $\tilde{B}^{(0)}_{z}$,which would be determined by the power of laser source. In order to realize the resonant response between the EM and the HFGWs, it is prerequisite to make sure $k_{e}=k_{g}$ as possible. The impact of bandwidth will be discussed in the following context.

Unlike pure plane EM waves, the EM components of GB are not completely perpendicular to the propagation direction (z-axis) except for the plane where the beam waist exists. Therefore, $\tilde{B}^{(0)}_{z}$ of GB would couple to $\tilde{E}^{(1)}_{y}$ and generate the x-component of the first-order perturbative EM power flux density (i.e., transverse perturbative photon flux $N^{(1)}_{x}$) . That is why we focus the major power of Gaussian beam on $\tilde{B}^{(0)}_{z}$ to enhance the perturbative transverse energy flux. Due to the x-component of the first-order perturbative EM power flux density including the effect of HFGWs, it can be considered as signal. We discuss the corresponding photon number (i.e., perturbative photon flux $N^{(1)}_{x}$ (PPFs)). The density of PPFs is
\begin{equation}
\label{PPFd}
n^{(1)}_{x}=\frac{1}{\hbar\omega_{e}}\left<\frac{1}{\mu_{0}}(\tilde{E}^{(1)*}_{y}\cdot\tilde{B}^{(0)}_{z})\right>,
\end{equation}
where $<>$ denotes the average value over a period. For a certain receiving area (on $y-z$ plane)
\begin{equation}
\label{PPF}
N^{(1)}_{x}=\int_{z}\int_{y}n^{(1)}_{x}dydz.
\end{equation}

 \rm{Relic HFGWs:}  the PPFs should be averaged by $\delta$ since the stochastic initial phase, \begin{equation}
\label{PPFrelic}
N^{(1)}_{x}=\frac{1}{2\pi}\int^{2\pi}_{0}d\delta\int_{z}\int_{y}n^{(1)}_{x}dydz.
\end{equation}

\rm{HFGWs from braneworld:}
\begin{equation}
\label{PPFbrane}
N^{(1)}_{x}=\int_{z}\int_{y}n^{(1)}_{x}dydz.
\end{equation}
According to the Eq.(\ref{perRelic})$-$ (\ref{PPFbrane}), the electromagnetic response of the Relic HFGWs and the braneworld HFGWs can be determined. Fig.2 illuminates the properties of PPFs from Relic HFGWs and braneworld HFGWs, which reveals (1) for each kind of HFGWs, the PPFs behaves as an even function on the $x$ axis, and the maximum PPFs always appear at $x=0$ plane with $z>0$. (2)Since the amplitude of braneworld HFGWs exceeds to the Relic HFGWs' amplitude $4$ orders, the PPFs from braneworld HFGWs always stronger the same order than the results of Relic HFGWs. Therefore, the PPFs on $x=0$ plane toward to positive $z$ direction is our ideal signal.

\begin{figure}[htbp]
\label{figNxz}
\centerline{\includegraphics[height=3.5cm]{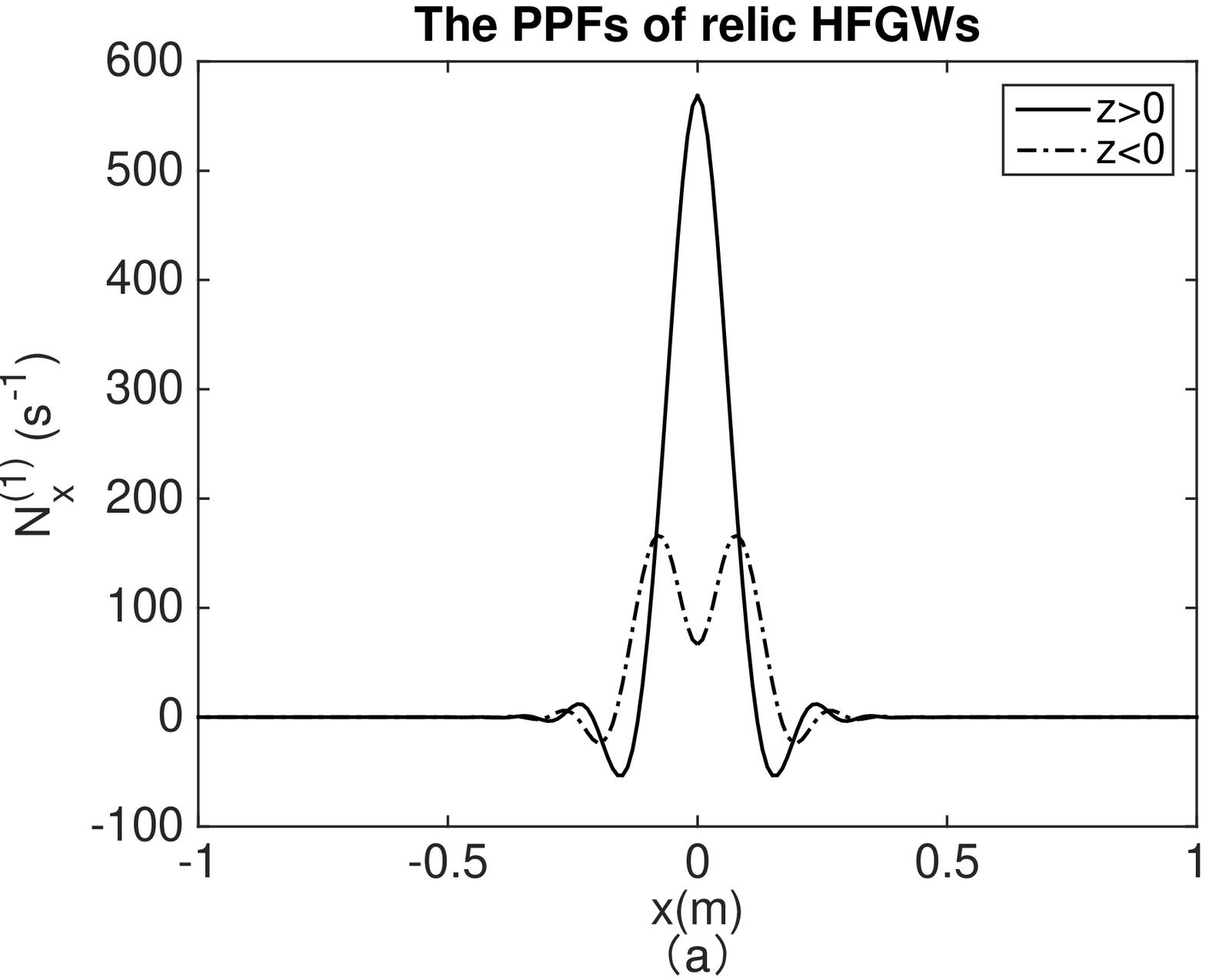}\includegraphics[height=3.5cm]{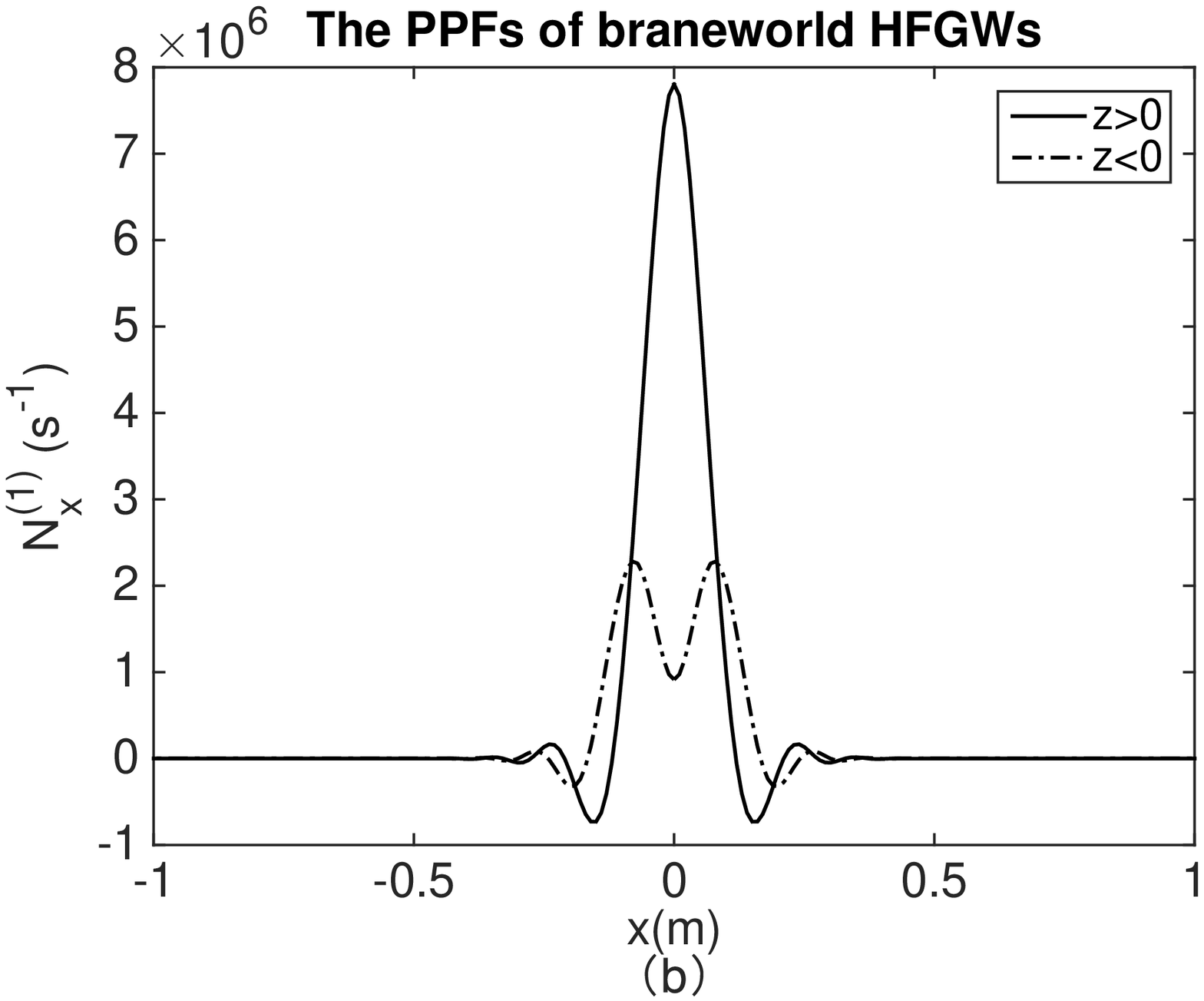}}
\centerline{\includegraphics[height=3.5cm]{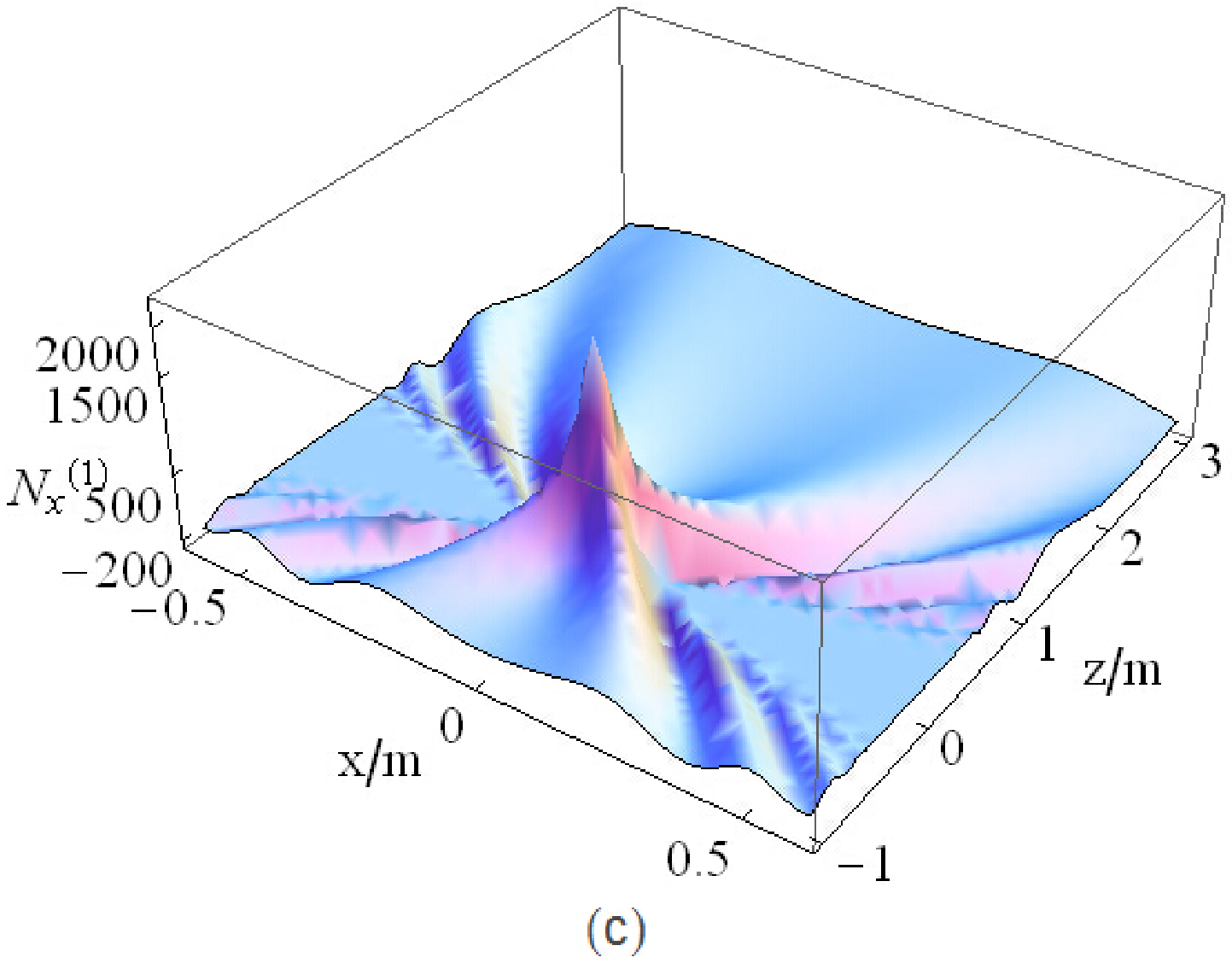}\includegraphics[height=3.5cm]{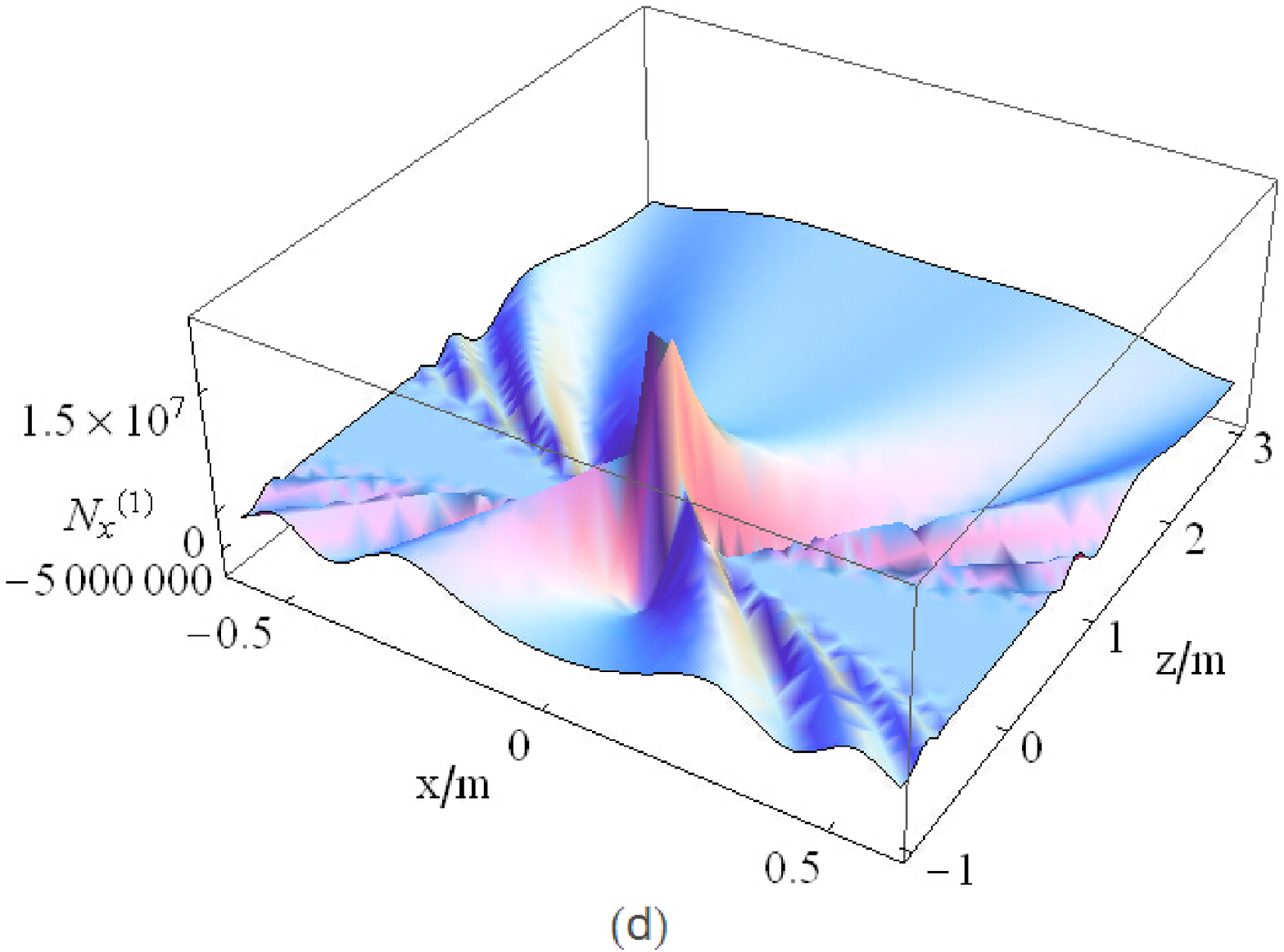}}
\caption{ (a) The perturbative photon fluxes of relic HFGWs, assuming $A_{\oplus}=A_{\otimes}=10^{-30}$; (b) The perturbative photon fluxes of HFGWs from braneworld with $A_{\oplus}=A_{\otimes}=10^{-26}$ and setting $\delta=1.23\pi$; (c) The 3D plot of the corresponding transverse perturbation photon fluxes $N_{x}^{(1)}$ produced by the Relic HFGWs; (d) The 3D plot of the corresponding transverse perturbation photon fluxes $N_{x}^{(1)}$ from braneworld HFGWs. }
\end{figure}

In the resonant system the GB as background supplies a longitudinal magnetic field component $\tilde{B}^{(0)}_{z}$, to produce the signal (PPFs). However it also causes the major noise of the system called as the background photon fluxes (BPFs). That can be derived as
\begin{equation}
\label{BPFd}
n^{(0)}_{x}=\frac{1}{\hbar\omega_{e}}\left<\frac{1}{\mu_{0}}(\tilde{E}^{(0)*}_{y}\cdot\tilde{B}^{(0)}_{z})\right>,
\end{equation}
and
\begin{equation}
\label{BPF}
N^{(0)}_{x}=\int_{z}\int_{y}n^{(0)}_{x}dydz.
\end{equation}
It is worthy to emphasize that the difference between the PPFs and BPFs comes from the different coupling electric components. It can be found the BPFs is odd symmetry for $x=0$ and when $z>0$ they propagates depart from $x=0$ (outgoing), while when $z<0$ they face to each other (imploding), and $N^{(0)}_{x}|_{x=0}=0$, where PPFs reach up to maximum. That implies on the $x=0$ plane the signal-to-noise ratio can reach considerable value so that in this place the detection will be realized in a suitable way. In practice this optimal plane is extremely narrow, so a new material or equipment which can play a focusable role in the system is very urgent.

According to the requirement of signal detection, the detection sensitivity in such case can be estimated as follows,
\begin{equation}
|N^{(1)}_{x}|^{2}\Delta t_{total}\geq|N^{noise}_{x}|,
\end{equation}
where $\Delta t_{total}$ is the requisite total signal accumulation time, and $N^{noise}_{x}$ includes the noisy photons mainly from BPFs. If we restrict the $\Delta t_{total}=3\rm{months}$, the most conservative estimation (i.e.,the major background noise (BPFs) with maximum value $\sim6\times 10^{22}\rm{s}^{-1}$, the dimensionless amplitude of the HFGWs should reach up to $A_{\oplus}(A_{\otimes})\sim 10^{-26}$. That means without any other additional equipment, the HFGWs from braneworld might be detectable in this scheme. However for the relic HFGWs, the system need to employ some instrument having signal focusing performance e.g. the fractal membranes\cite{epjc57}-\cite{epjc59}.

In order to maintain the maximum of PPFs in a wider transverse distance,  we use the fractal membranes (FM) with their plane overlap the $x=0$ plane where $N^{(0)}_{x}=0$ and $N^{(1)}_{x}$ has its maximum value (cf.fig.1), which can keep the maximum strength of the PPFs ($N^{(1)}_{x}$) invariant within certain distance\cite{epjc57},\cite{epjc58}. As our previous works \cite{EPJC},\cite{grg}, it is considered to be that after $N^{(1)}_{x}$ is transmitted, it can keep its $90$ percent strength in 1 meter. However there is no BPFs on FM plane, so FM is inutility for BPFs. Table I lists the BPFs and PPFs after using the FM. In the region of $(66, 70]\rm{cm}$, the EM effect from relic GWs would be detectable.
\begin{table}
\label{FM}
\caption{Comparison of BPFs ($|N^{(0)}_{x}|$) and PPFs
($|N^{(1)}_{x}|_{\rm{FM}}$) with fractal membrane in the x-direction. Note beyond $0.7$m the photon number of the BPFs is less than 1, so beyond the area it can be neglected.}
\begin{center}
\begin{tabular}{llllll}
\hline\noalign{\smallskip}
$x(m)$  & 0 & 0.05 & 0.61 & 0.66 & 0.70 \\
\hline\noalign{\smallskip}\noalign{\smallskip}
$|N_{x}^{(0)}|(s^{-1})$ & 0 & $5.8\times10^{22}$ & $4.6\times10^{5}$ & 603 & 2\\
$\rm{relic}: |N_{x}^{(1)}|_{FM}(s^{-1})$ & 568.40 & 565.60 & 533.70 & 530.90 & 528.60\\
$\rm{braneworld}: |N_{x}^{(1)}|_{FM}(s^{-1})$ & 7.80$\times10^{6}$ & 7.76$\times10^{6}$ & 7.32$\times10^{6}$ & 7.28$\times10^{6}$ & 7.25$\times10^{6}$\\
\noalign{\smallskip}\hline
\end{tabular}
\end{center}
\end{table}

In experiment, it is always expected to obtain the optimal effect. Thus how to select parameters is another issue for us, following the principle of maximum SNR we discuss the optimal system parameters, such as waist of GB ($W_{0}$), bandwidth and resonance frequency ($\nu$).

Firstly, the waist of GB ($W_{0}$) plays significant role on the transverse PPFs production. Fig.3 describes the background photon fluxes (BPFs) and the perturbative photon fluxes (PPFs) varying different waist of GB. That indicates the larger waist of GB ($W_{0}$) can result in stronger signal energy, and weaken the strength of the major background noise BPFs. Therefore, for the given conditions ($\nu=3\rm{GHz}, \hat{B}_{y}^{(0)}=6\rm{T}, P=10\rm{W}$), the waist $W_{0}=9\rm{cm}$ could achieve the highest SNR.
\begin{figure}[htbp]
\centerline{\includegraphics[height=4.5cm]{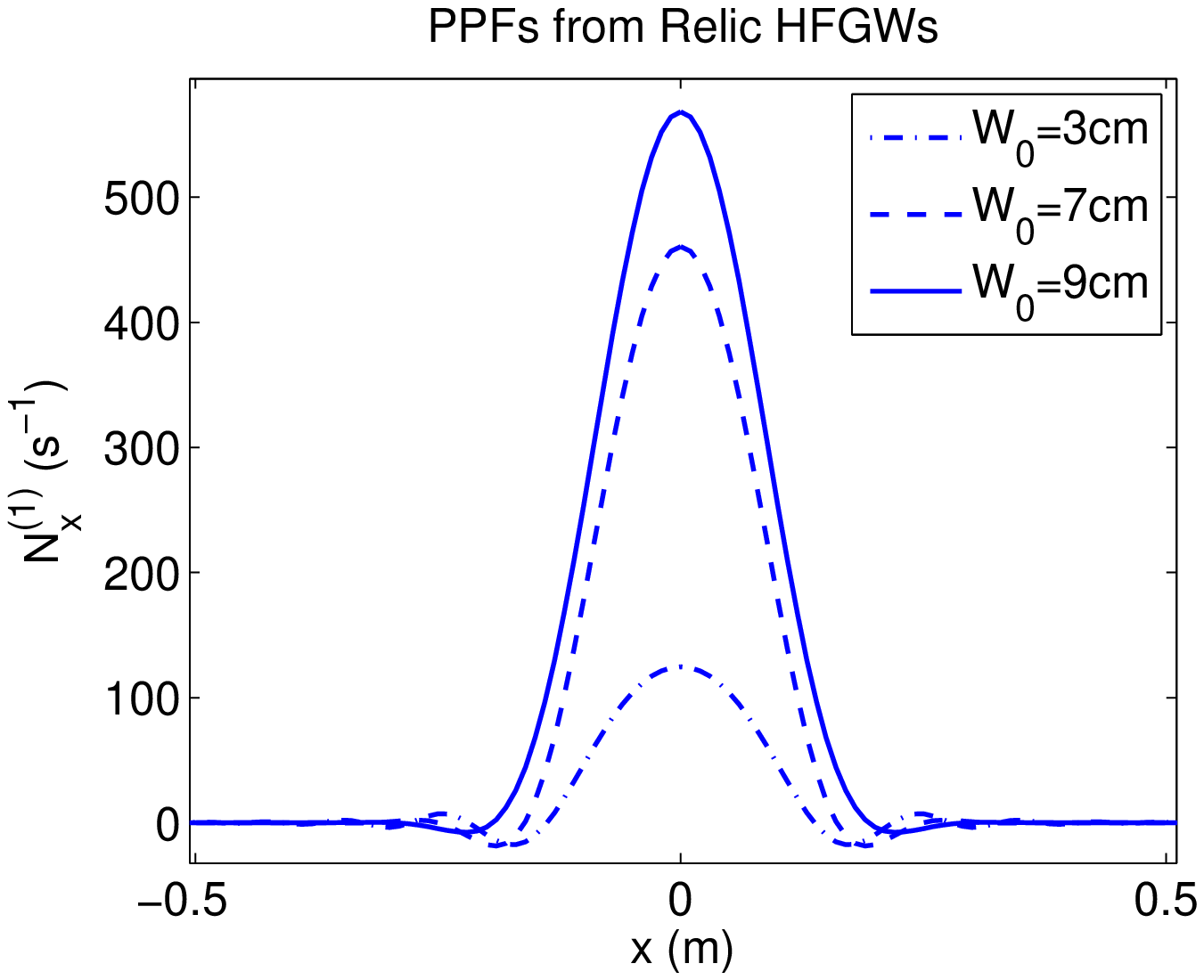}\includegraphics[height=4.5cm]{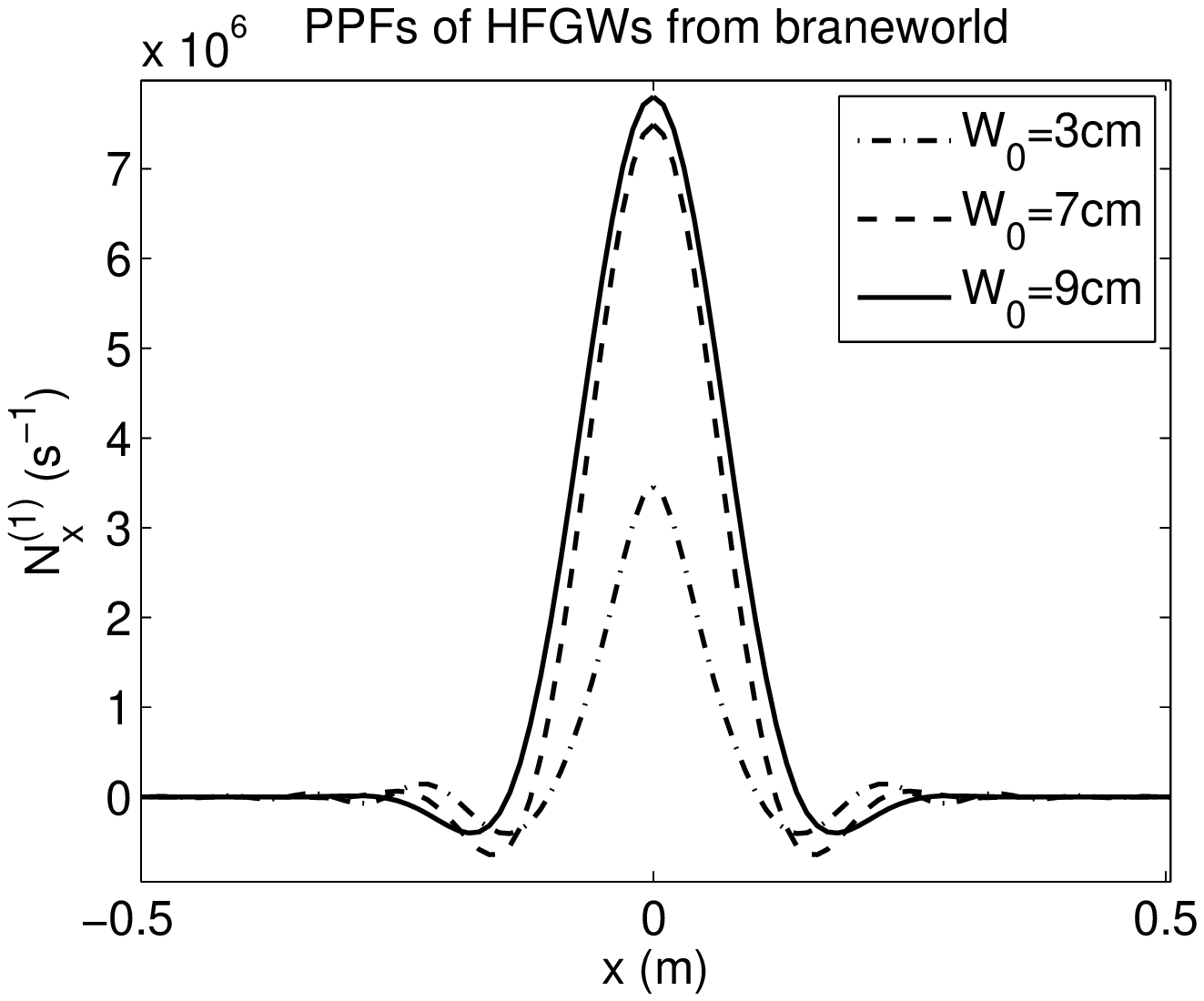}\includegraphics[height=4.5cm]{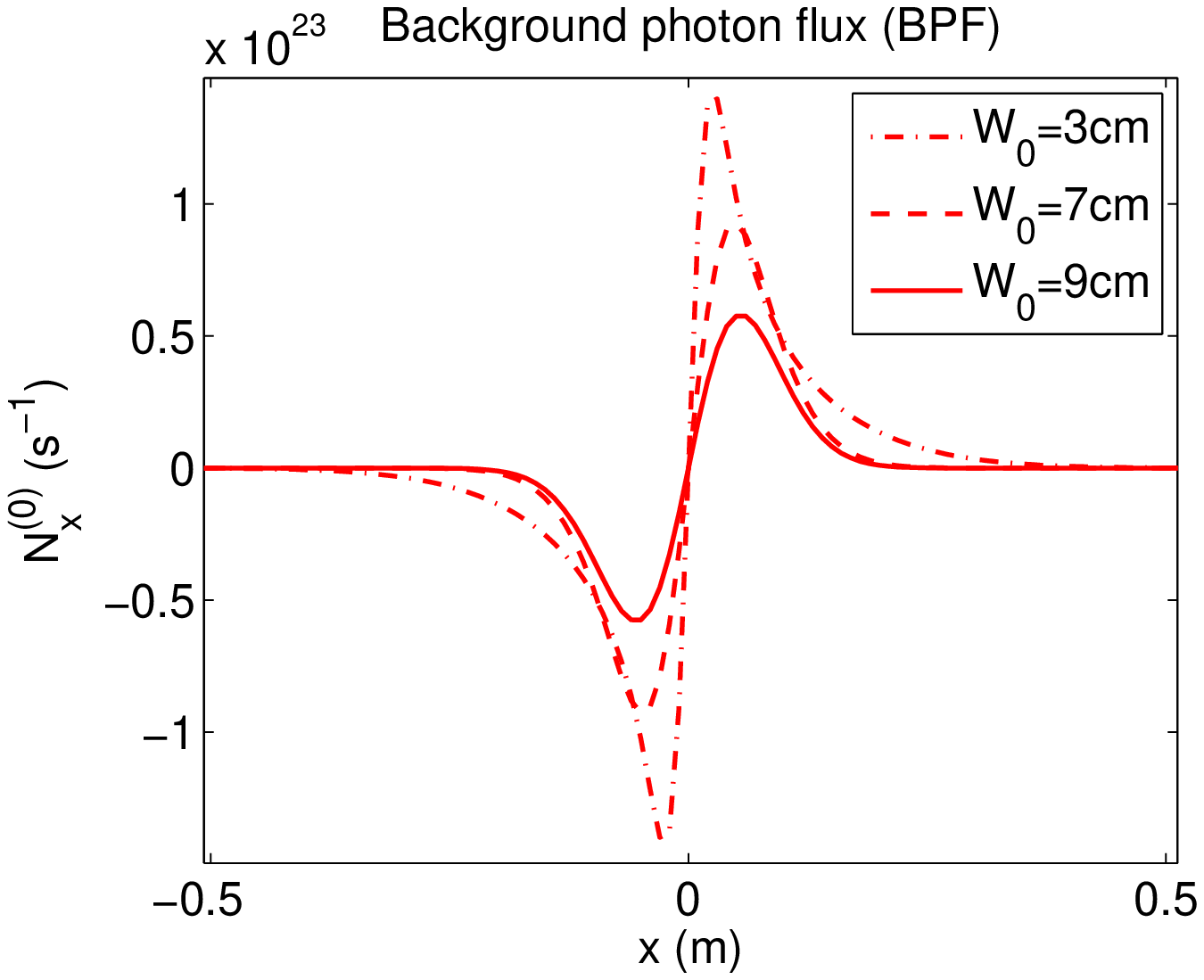}}
\label{figw0} \caption{The effect of the GB waist on the transverse background photon fluxes (BPFs) and perturbation photon fluxes (PPFs). Here $\nu=3\rm{GHz},\hat{B}_{y}^{(0)}=6\rm{T}, P=10\rm{W},z>0$ and for braneworld HFGWs supposing $\delta=1.23\pi$.}
\end{figure}

Secondly, the bandwidth of the laser is inevitable. Now about the effect of bandwidth in our system is also taken into our concern. According to Eq.(\ref{PPFd}), the PPFs should be proportional to $\langle \rm{cos}(k_{g}z-\omega_{g}t)\rm{cos}(k_{e}z-\omega_{e}t+\delta)\rangle$, which can be separated into the following two intergrations,
\ba &&
\tilde{n}^{(1)}_{x}\propto\left(\omega_{g}-\omega_{e}\right)\int^{\frac{2\pi}{\omega_{g}-\omega_{e}}}_{0}\rm{cos}\left((k_{g}-k_{e})z-(\omega_{g}-\omega_{e})t-\delta\right)dt+
\nonumber\\&&
\left(\omega_{g}+\omega_{e}\right)\int^{\frac{2\pi}{\omega_{g}+\omega_{e}}}_{0}\rm{cos}\left((k_{g}+k_{e})z-(\omega_{g}+\omega_{e})t+\delta\right)dt.
\ea
So when $\omega_{g}=\omega_{e}$, $\tilde{n}^{(1)}_{x}\propto\rm{cos}\delta$, while $\omega_{g}\neq\omega_{e}$, $\tilde{n}^{(1)}_{x}=0$. That means the components of GWs and GB with different frequency can not resonate to each other. So the frequency bandwidth would not influence the original signal.
\begin{figure}[htbp]
\centerline{\includegraphics[height=4.5cm]{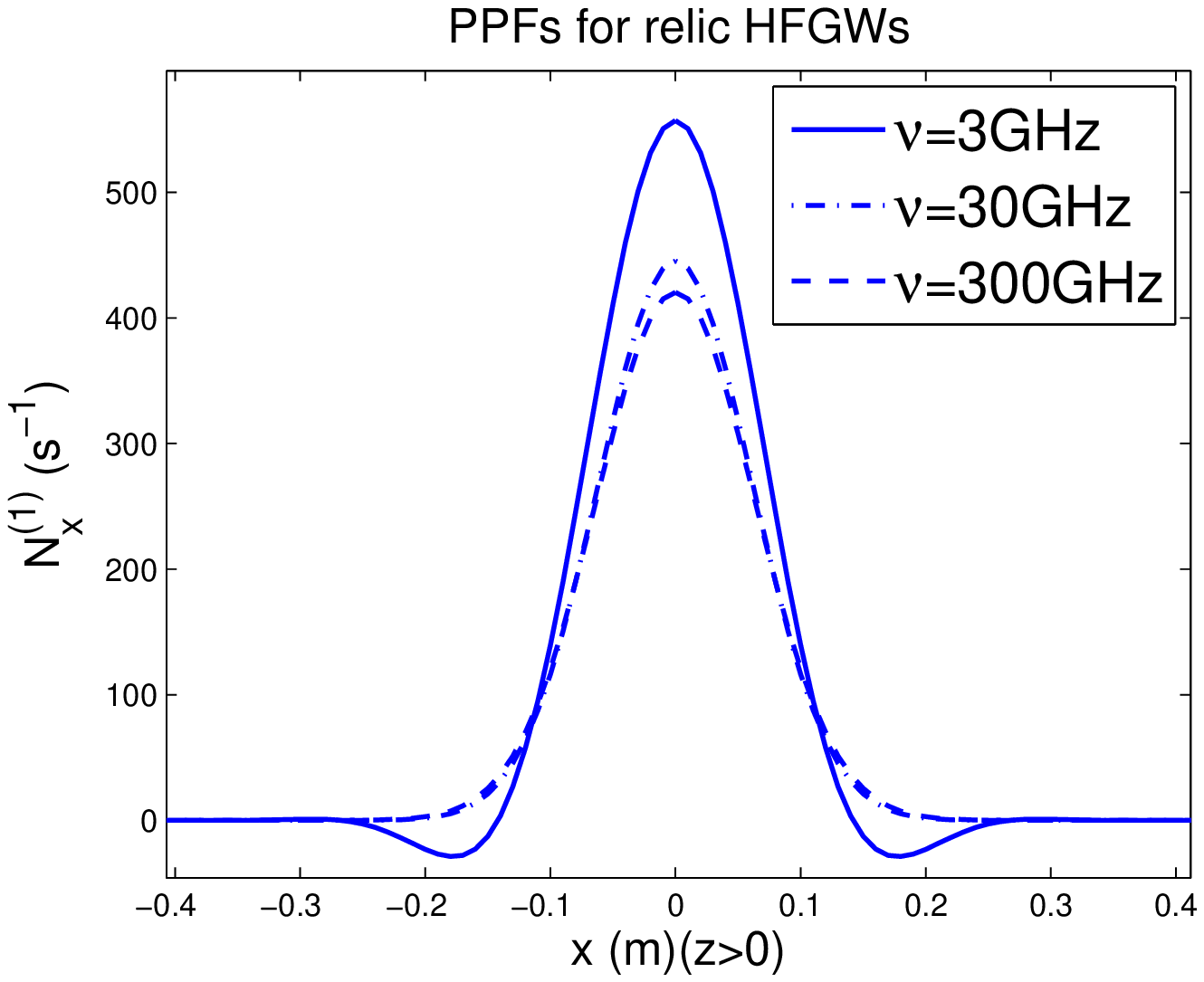}\includegraphics[height=4.5cm]{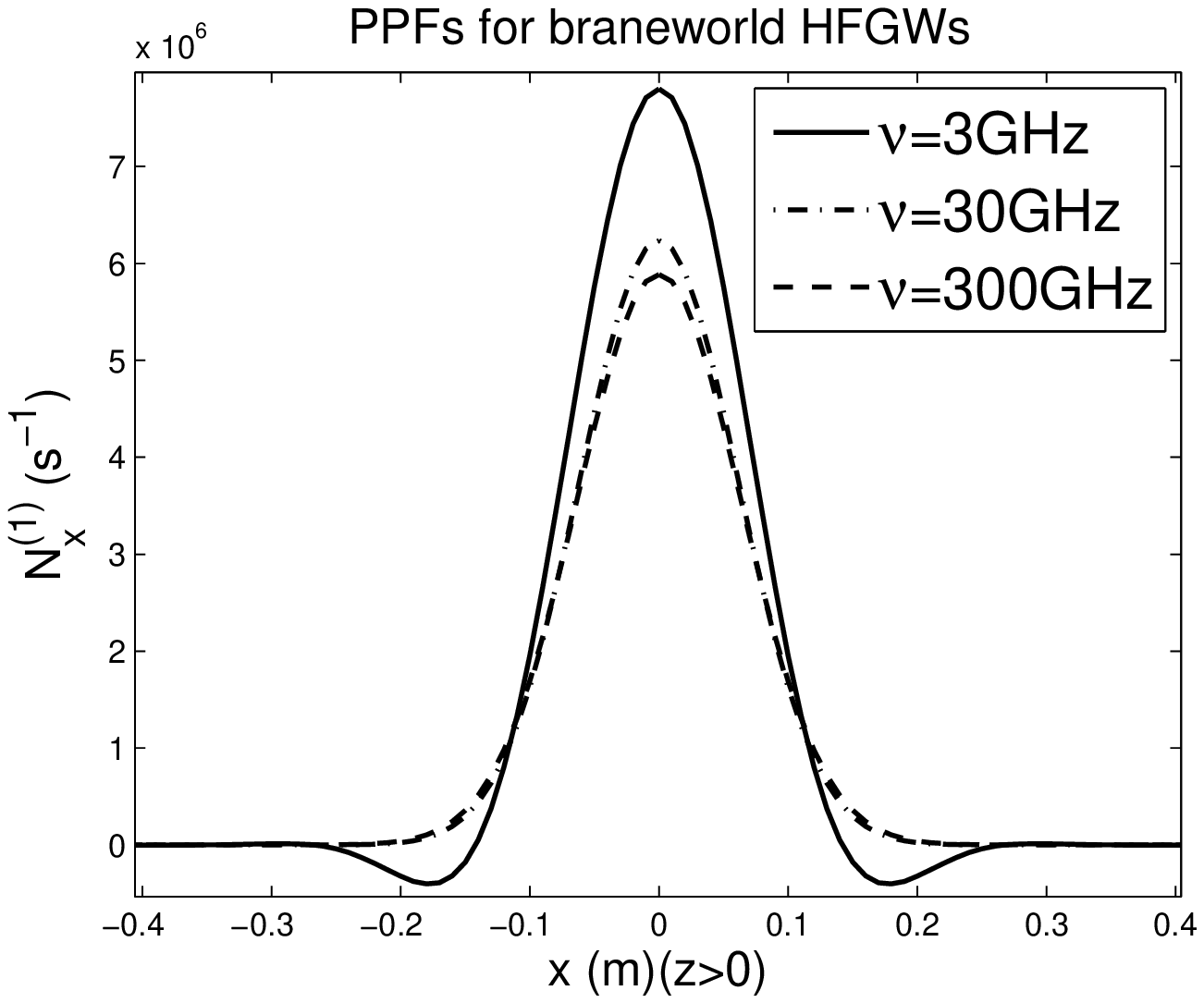}\includegraphics[height=4.5cm]{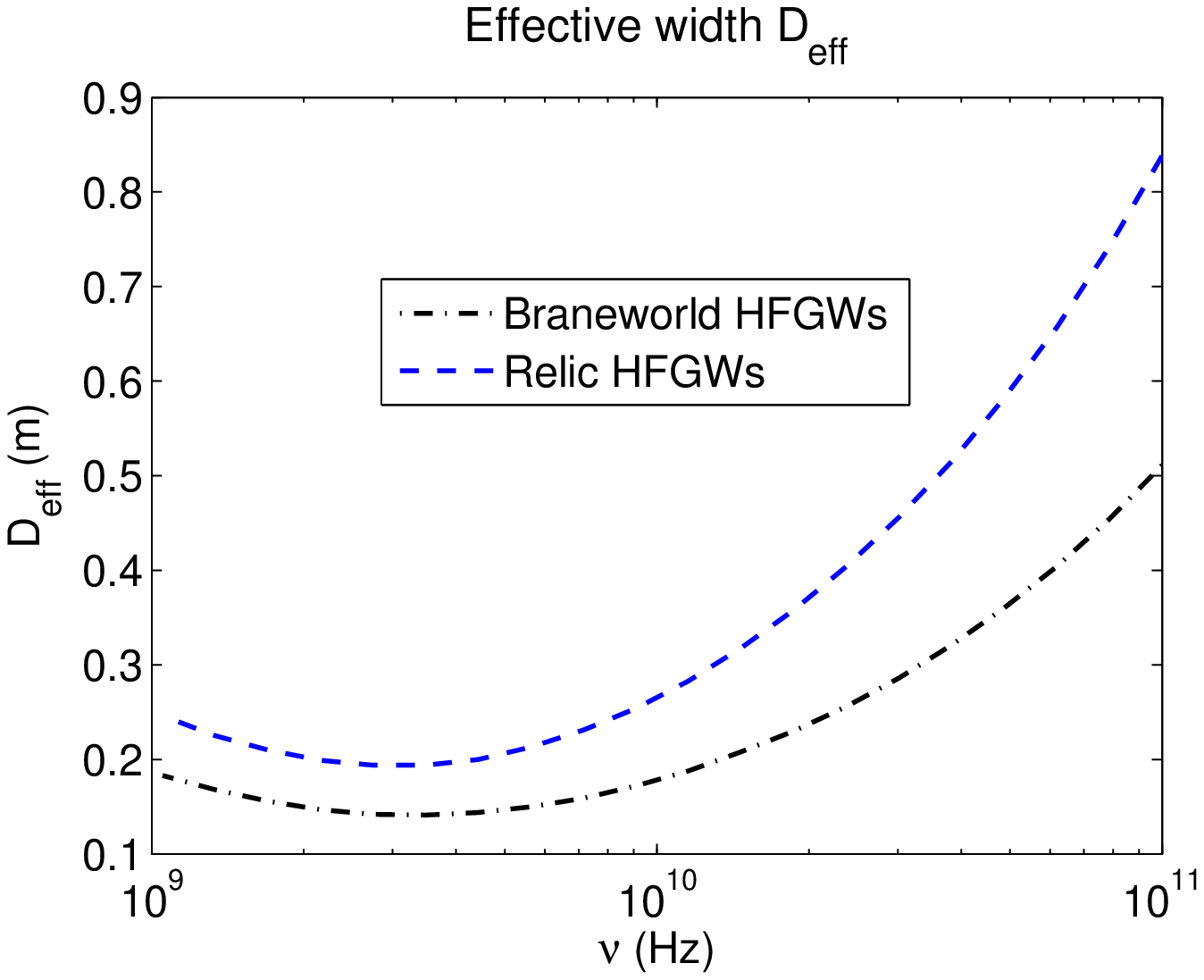}}
\label{fig1} \caption{The effect of frequency. Here choosing the waist as the optimal value $W_{0}=9\rm{cm}$.}
\end{figure}

Finally, it should be clarified that theoretically the frequency of the HFGWs can range from $10^{8}$Hz to $10^{10}$Hz for the relic HFGWs and even reach up to $10^{14}$Hz in braneworld case\cite{extrabrane2}. Thus the resonance frequency $\nu$ should be variable. The corresponding results in Fig.4 illuminates that with the frequency increase, the maximum values of PPFs will decrease slightly, but the directive property of PPFs becomes much better(i.e. the propagation direction of signal energy fluxes tends to be non-oscillating state). So it is meaningful to define the transverse area around $x=0$ plane with positive PPFs as effective width $D_{\rm{eff}}$. Then Fig.4 has shown the general relationship between the effective width and frequency. It can be found with the frequency growth the effective width will become wider, in other words, in the higher frequency band the effective energy fluxes can be enhanced greatly. Furthermore it is worthy to emphasize that with the frequency increase the BPFs will decrease more quickly than the PPFs (cf. Table II), which means the SNR will be further improved in higher frequency band.
\begin{table}
\label{FM}
\caption{The maximum value of the PPFs ($N_{x}^{(1)}|_{\rm{max}}$) and BPFs ($N_{x}^{(0)}|_{\rm{max}}$) varying with some typical frequencies. For relic HFGWs we suppose $A_{\oplus}=A_{\otimes}=10^{-30}$. Although the dimensionless amplitude $h$ varies with frequency in ordinary inflation model, it can be held at this order of magnitude through choosing different inflation parameters~\cite{zhang2}. For braneworld HFGWs, assuming $A_{\oplus}=A_{\otimes}=10^{-26}$.}
\begin{center}
\begin{tabular}{lll}
\hline\noalign{\smallskip}
$\nu$(GHz)  &~~~~~~~~~~~~~~~~~~$N_{x}^{(1)}|_{\rm{max}}$~($s^{-1}$) & $N_{x}^{(0)}|_{\rm{max}}$~($s^{-1}$) \\
           &relic HFGWs~~~~~~~~braneworld HFGWs&          \\
\hline\noalign{\smallskip}\noalign{\smallskip}
3 &~~557.30~~~~~~~~~~~~~~~~~~~~~$7.80\times10^{6}$ &~~$1.40\times10^{23}$ \\
30&~~445.50~~~~~~~~~~~~~~~~~~~~~$6.24\times10^{6}$ &~~$5.03\times10^{21}$ \\
300&~~420.40~~~~~~~~~~~~~~~~~~~~~$5.89\times10^{6}$ &~~$7.14\times10^{18}$  \\
\noalign{\smallskip}\hline
\end{tabular}
\end{center}
\end{table}

As a conclusion, based on the GB resonant system of HFGWs, we discussed the EM effects from the relic HFGWs and the braneworld HFGWs contrastly, and found the optimal parameters and effective width to improve sensitivity. The main remarks of this paper can be classified as follows, $\rm{(1)}$ For the relic HFGWs (around GHz band) predicted by QIM may store a large amount of energy~\cite{QIM1,QIM2,QIM3,QIM4,QIM5}, the scheme discussed in this paper can be considered as a hopeful candidate for such GWs detection. Under the typical conditions of resonant response and consider the isotropic property of the relic GWs, the first-order transverse perturbative photon fluxes (PPFs) would be expected to be $\sim 5\times10^{2}\rm{s^{-1}}$, This result is much better than our previous work~\cite{grg} because the longitudinal background magnetic component is set to be the standard Gaussian form. Meanwhile as a coherent GWs source, the HFGWs from braneworld would generate much stronger PPFs $\sim7\times10^{6}\rm{s^{-1}}$ due to its quite larger dimensionless amplitude $h$. So if the braneworld scenarios is correct, our EM response system would be much more hopeful to capture the GWs from extra dimensional braneworld. $\rm{(2)}$ The system parameters play very important roles in signal detection. Focusing on the GB waist, frequency bandwidth and response frequency, we analyzed their effect on the SNR, and found the optimal waist $W_{0}=9\rm{cm}$ under the typical condition. Additionally in the possible HFGWs' frequency band, the higher response frequency would result in better detection effect (higher SNR and larger effective width for energy fluxes accumulation). The effects are independent from frequency bandwidth.

Based on above results, this paper presents new features to indicate possible observational effects caused by HFGWs predicted by the inflationary or braneworld model, in the proposed detection system. If these enhanced signals could be captured in future, it will involve meaningful issues not only about the direct detection of GWs, but also the very early universe, dimension of space and the multiverse. If not, these models or scenarios would need some corrections, or other system parameters require adjustment for further enhanced sensitivity by very different physical behaviors between the signals and background noise, such as different propagating directions, distributions, decay rates, wave impedance, and so on. These topics would be addressed in details as subsequent works.
\\
\\
\section*{\bf Acknowledgements}
This work is supported by FAPESP No. 2012/08934-0, CNPq, CAPES,National Natural Science Foundation of China No. 11205254, No. 11178018, No.11573022 and No.11375279, and the Fundamental Research Funds for the Central Universities 106112015CDJRC131216 and CDJRC10300003, and Chongqing Postdoctoral Science Foundation (No.Xm2015027), and the Open Project Program of State Key Laboratory of Theoretical Physics, Institute of Theoretical Physics, Chinese Academy of Sciences, China (No.Y5KF181CJ1).

\end{document}